# Phase-controlled transparent superluminal light propagation in a Doppler-broadened four-level N-type system

Lida Ebrahimi Zohravi[1], Majid Abedi[2] and Mohammad Mahmoudi[1]

[1] Department of Physics, University of Zanjan, University Blvd, 45371-38791, Zanjan, Iran
[2] Institute for Advanced Studies in Basic Sciences, P.O. Box 45195-159, Zanjan, Iran
E-mail: mahmoudi@znu.ac.ir

**Abstract.** The propagation of a weak probe field in a four-level N-type quantum system in the presence of spontaneously generated coherence (SGC) is theoretically investigated. The optical properties of the system are studied and it is shown that the group velocity of light pulse can be controlled by relative phase of applied fields. By changing the relative phase of applied fields, the group velocity of light pulse changes from transparent subluminal to the transparent superluminal light propagation. Thus, the phase-controlled absorption-free superluminal light propagation is obtained without applying an incoherent laser fields to the system. The propagation of a weak probe light pulse is studied by solving the Maxwell's wave equation on numerical grid in space and time. Moreover, we study the third order self- and cross-Kerr susceptibility of probe field and calculate the nonlinear cross-phase shift for different values of intensity of applied fields. In addition, we take into account the effect of Doppler broadening on the light pulse propagation and it is found that a suitable choice of laser propagation directions allows us to preserve our results even in the presence of Doppler effect. It is demonstrated that by increasing the Doppler width of distribution to the room temperature, the dispersion changes from transparent subluminal to transparent superluminal light propagation which is our major motivation for this work.

## I. INTRODUCTION

Atomic coherence plays a key role in modification and controlling the optical properties of quantum systems in atom-field interaction. It can be generated by applying the coherent laser fields to the quantum systems or by quantum interference due to the spontaneous emission. One of the interesting applications of atomic coherence is that it modifies light pulse propagation through the dispersive medium such as slow or fast light which it has a major role in implementation of quantum information processing and in all-optical communication system.

The original idea of group velocity was introduced by W. R. Hamilton at 1839 [1]. Lord Rayleigh proposed the group velocity concept instead of phase velocity for propagation of light pulse inside a medium at the end of 19th century and then developed the theory of anomalous dispersion [2]. In his study the group, energy and signal velocities have used as unrecognizable concept. The problem he faced, was the theory of relativity [3], however it was subsequently solved by Sommerfeld and Brillouin [4].

It is well known that the velocity of pulse peak in an anomalous dispersive medium can exceed the speed of light in vacuum, leading to the superluminal light propagation. The phenomenon of propagation with negative group velocity was theoretically predicted by Garrett and McCumber [5] and experimentally was first performed by Chu and co-workers using a correlation technique [6]. However, such phenomena cannot transmit the information with a velocity faster than the speed of light in vacuum and it does not violate Einstein's special theory of relativity [7-9].

Experimental and theoretical studies on controlling the group velocity of light pulse inside different material has been extensively studied [10-13]. It was demonstrated that switching from subluminal to superluminal pulse propagation can be controlled by the intensity or relative phase of applied fields [14-17]. The quantum interference due to the spontaneous emission, i.e. spontaneously generated coherence (SGC) [18] is another technique for controlling the optical properties of systems [19-21].

The gain-assisted superluminal light propagation was first observed experimentally by Wang et al [12]. It was also shown that such light propagation can be obtained in coupled optical resonators [22], the four-level quantum system [23] and in five-level dressed interacting ground-state system [24]. An ideal condition for light propagation is a region in which the system does not show the absorption or gain of the probe field. This is due to the fact that the large absorption in the system does not permit the pulse propagates inside the medium. On the other hand, the gain may add some noise to the system.

The optical properties of N-type four-level quantum system have been extensively studied. The theoretical and experimental results of researches were reported in several papers [25-28]. The stimulated emission and multi-peaked absorption in a four-level N-type atomic system were also investigated and it was shown that in the absence of SGC, changing the intensity of applied fields switches the slope of dispersion from positive to negative [29]. The effect of SGC on the optical properties of N-type system was also investigated [30-32]. Recently the phase control of probe response in a Doppler-broadened four-level N-type system in the presence of SGC was reported [33]. They have shown that in the presence of SGC, the optical properties of such system are phase-dependent. In this paper, we investigate the optical light pulse propagation through a four-level N-type system and we show that, by choosing the suitable set of parameters, the gain-assisted transparent superluminal light propagation can be established in this system.

The possibility of such propagation is investigated by numerical solving the Maxwell's wave equation for a probe pulse field. In addition, we investigate the effect of the third order nonlinear susceptibility on the nonlinear phase shift of a probe field. Moreover, it is demonstrated that by taking into account the effect of SGC, the slope of dispersion can be changed from positive to negative, just by changing the relative phase of applied fields which is another important motivation of this work. Finally, we include the effect of Doppler broadening on the light pulse propagation and it is shown that by increasing the Doppler width of distribution to the room temperature, the dispersion changes from transparent subluminal to transparent superluminal light propagation. Our results can be preserved, even in the presence of Doppler effect, by the same propagation direction of applied fields (co propagating).

## II. MODEL AND EQUATIONS

Our interesting system is a four-level $N$-type atomic system which is shown in figure 1. In our study, we consider four dipole-allowed transitions: $|1> - |3>$, $|2> - |3>$, $|1> - |4>$ and $|2> - |4>$. The direct transitions $|1> - |2>$ and $|3> - |4>$ are dipole forbidden. The transition $|2> - |3>$ of frequency $\omega_{32}$ is coupled by a coupling field $[\boldsymbol{\varepsilon_c} = \mathbf{E}_c \exp(-i\omega_c t) + c.c.]$ with Rabi frequency $\Omega_c = \mu_{32}.E_c/(2\hbar) = |\Omega_c|\exp(i\phi_c)$. A coherent pumping field $[\boldsymbol{\varepsilon_s} = \mathbf{E}_s \exp(-i\omega_s t) + c.c.]$ with real Rabi frequency $\Omega_s = \boldsymbol{\mu}_{41}.\mathbf{E}_s/(2\hbar)$ excites the transition $|1> - |4>$. The transition $|1> - |3>$ of frequency $\omega_{31}$ is driven by a weak probe field $[\boldsymbol{\varepsilon_p} = \mathbf{E}_p \exp(-i\omega_p t) + c.c.]$ with Rabi frequency $\Omega_p = \mu_{31}.E_p/(2\hbar) = |\Omega_p|\exp(i\phi_p)$. In our notation, the parameter $\boldsymbol{\mu}_{ij} = \langle i| -e\mathbf{r}|j\rangle$ is the dipole matrix element of $|i> - |j>$ transition.

As a realistic example, we consider hyperfine levels of $^{85}Rb$ which is shown in figure 1. The transition $5S_{1/2}, F = 2 \rightarrow 5P_{1/2}, F^{'} = 3$ is coupled by a probe field. The coupling field couples the transition $5S_{1/2}, F = 3 \rightarrow 5P_{1/2}, F^{'} = 3$. Both transitions can be excited by a diode laser at the frequency $795\,nm$. The pumping field also couples the transition $5S_{1/2}, F = 2 \rightarrow 5P_{3/2}, F^{'} = 4$ which can be provided by a Ti: sapphire laser [31].

The Hamiltonian of the system in dipole and rotating wave approximation is given by:

$$H = \Omega_s \exp(i\Delta_s t)|4><1| + \Omega_p \exp(i\Delta_p t)|3><1| + \Omega_c \exp(i\Delta_c t)|3><2| + H.c. \,, \qquad (1)$$

where $\Delta_s = \omega_s - \omega_{41}$, $\Delta_c = \omega_c - \omega_{32}$ and $\Delta_p = \omega_p - \omega_{31}$ are the detuning of pumping, coupling and probe fields, respectively.

Using the Von Neumann equation for density matrix, $i\hbar\dot{\rho} = [H,\rho]$, the equations of motion in the rotating-wave approximation and in rotating frame are given by

$$\dot{\tilde{\rho}}_{11} = \gamma_{31}\tilde{\rho}_{33} + \gamma_{41}\tilde{\rho}_{44} + i|\Omega_p|\tilde{\rho}_{31} - i|\Omega_p|\tilde{\rho}_{13} + i\Omega_s\tilde{\rho}_{41} - i\Omega_s\tilde{\rho}_{14},$$

$$\dot{\tilde{\rho}}_{22} = \gamma_{32}\tilde{\rho}_{33} + \gamma_{42}\tilde{\rho}_{44} + i|\Omega_c|\tilde{\rho}_{32} - i|\Omega_c|\tilde{\rho}_{23},$$

$$\dot{\tilde{\rho}}_{33} = -(\gamma_{31} + \gamma_{32})\tilde{\rho}_{33} + i|\Omega_p|\tilde{\rho}_{13} - i|\Omega_p|\tilde{\rho}_{31} + i|\Omega_c|\tilde{\rho}_{23} - i|\Omega_c|\tilde{\rho}_{32},$$

$$\dot{\tilde{\rho}}_{44} = -(\gamma_{41} + \gamma_{42})\tilde{\rho}_{44} + i\Omega_s\tilde{\rho}_{14} - i\Omega_s\tilde{\rho}_{41},$$

$$\dot{\tilde{\rho}}_{12} = -i(\Delta_p - \Delta_c)\tilde{\rho}_{12} + i|\Omega_p|\tilde{\rho}_{32} + i\Omega_s\tilde{\rho}_{42} - i|\Omega_c|\tilde{\rho}_{13} + \sqrt{\gamma_{31}\gamma_{32}}\,p_1\eta_1\tilde{\rho}_{33}\exp(i\phi) +$$
$$\sqrt{\gamma_{41}\gamma_{42}}\,p_2\eta_2\tilde{\rho}_{44}\exp(i\phi),$$

$$\dot{\tilde{\rho}}_{13} = -[(\gamma_{31} + \gamma_{32})/2 + i\Delta_p]\tilde{\rho}_{13} + i\Omega_s\tilde{\rho}_{43} - i|\Omega_c|\tilde{\rho}_{12} + i|\Omega_p|(\tilde{\rho}_{33} - \tilde{\rho}_{11}),$$

$$\dot{\tilde{\rho}}_{14} = -[(\gamma_{41} + \gamma_{42})/2 + i\Delta_s]\tilde{\rho}_{14} + i|\Omega_p|\tilde{\rho}_{34} + i\Omega_s(\tilde{\rho}_{44} - \tilde{\rho}_{11}),$$

$$\dot{\tilde{\rho}}_{23} = -[(\gamma_{31} + \gamma_{32})/2 + i\Delta_c]\tilde{\rho}_{23} - i|\Omega_p|\tilde{\rho}_{21} + i|\Omega_c|(\tilde{\rho}_{33} - \tilde{\rho}_{22}),$$

$$\dot{\tilde{\rho}}_{24} = -[(\gamma_{41} + \gamma_{42})/2 + i(\Delta_s - \Delta_p + \Delta_c)]\tilde{\rho}_{24} + i|\Omega_c|\tilde{\rho}_{34} - i\Omega_s\tilde{\rho}_{21},$$

$$\dot{\tilde{\rho}}_{34} = -[(\gamma_{31} + \gamma_{32} + \gamma_{41} + \gamma_{42})/2 + i(\Delta_s - \Delta_p)]\tilde{\rho}_{34} + i|\Omega_p|\tilde{\rho}_{14} + i|\Omega_c|\tilde{\rho}_{24} - i\Omega_s\tilde{\rho}_{31}, \qquad (2)$$

where $\tilde{\rho}_{11} + \tilde{\rho}_{22} + \tilde{\rho}_{33} + \tilde{\rho}_{44} = 1$. The rotating frame, considered here, is defined by

$$\tilde{\rho}_{11} = \rho_{11}, \tilde{\rho}_{22} = \rho_{22}, \tilde{\rho}_{33} = \rho_{33}, \tilde{\rho}_{44} = \rho_{44}, \quad \tilde{\rho}_{12} = \rho_{12}e^{-i[(\Delta_c - \Delta_p)t + (\phi_c - \phi_p)]}, \quad \tilde{\rho}_{13} = \rho_{13}e^{i(\Delta_p t + \phi_p)},$$

$$\tilde{\rho}_{14} = -\rho_{14}e^{i\Delta_s t}, \quad \tilde{\rho}_{23} = -\rho_{23}e^{i(\Delta_c t + \phi_c)}, \quad \tilde{\rho}_{24} = -\rho_{24}e^{i[(\Delta_s + \Delta_c - \Delta_p)t + (\phi_c - \phi_p)]}, \quad \tilde{\rho}_{34} = \rho_{34}e^{i[(\Delta_s - \Delta_p)t - \phi_p]}. \qquad (3)$$

Here $\gamma_{mn}$ is the spontaneous decay rate from level $|m>$ to level $|n>$. The parameter $\phi = \phi_p - \phi_c$ shows the relative phase of Probe and coupling fields. The spontaneous decays from level $|2> (|4>)$ to level $|1> (|3>)$ is ignored. We assume that levels $|1> (|3>)$ and $|2> (|4>)$ are so closely that both of decay channels from upper level $|3> (|4>)$ to the lower levels are interacting with the same vacuum mode and then the SGC is established. The terms containing $\sqrt{\gamma_{31}\gamma_{32}}\,p_1\eta_1\tilde{\rho}_{33}$ and $\sqrt{\gamma_{41}\gamma_{42}}\,p_2\eta_2\tilde{\rho}_{44}$ show such quantum interference. Note that for nearly degenerate lower levels, i.e., $\omega_{31} \cong \omega_{32}$ $(\omega_{41} \cong \omega_{42})$, quantum coherence becomes important and then $\eta_1(\eta_2) = 1$. However for large values of lower levels spacing, the effect of quantum interference may be dropped and then $\eta_1(\eta_2) = 0$. The parameter $p_1 \equiv \boldsymbol{\mu}_{31}.\boldsymbol{\mu}_{32}/|\boldsymbol{\mu}_{31}||\boldsymbol{\mu}_{32}|$ $(p_2 \equiv \boldsymbol{\mu}_{41}.\boldsymbol{\mu}_{42}/|\boldsymbol{\mu}_{41}||\boldsymbol{\mu}_{42}|)$ depends on the alignment of the two dipole moments $\boldsymbol{\mu}_{31}(\boldsymbol{\mu}_{41})$ and $\boldsymbol{\mu}_{32}(\boldsymbol{\mu}_{42})$. When the two dipole moments are parallel, the effect of quantum interference is maximum and $p_1(p_2) = 1$ whereas for the orthogonal dipole moments there is no interference due to spontaneous emission and $p_1(p_2) = 0$.

When $\Delta_c = \Delta_p = 0$, $\eta_1 = \eta_2 = \eta$, $p_1 = p_2 = p$ and $\gamma_{31} = \gamma_{32} = \gamma_{41} = \gamma_{42} = \gamma$, the analytical expressions for probe, coupling and pumping transition coherences in weak probe field approximation are given by

$$\tilde{\rho}_{31} = \frac{i\gamma(a_0 + a_1|\Omega_c|^2|\Omega_p|)}{d(d_0 + d_1\Delta_p + d_2\Delta_p^2 - 4i\gamma\Delta_p^3 + \Delta_p^4)}, \tag{4-a}$$

$$\tilde{\rho}_{32} = \frac{i\Omega_s^2|\Omega_c|}{d}, \tag{4-b}$$

$$\tilde{\rho}_{41} = \frac{i\gamma|\Omega_c|^2\Omega_s}{d}, \tag{4-c}$$

where

$$a_0 = 2p\eta e^{i\phi}|\Omega_c|^3\Omega_s^2(2\gamma^2 + |\Omega_c|^2 - \Omega_s^2 + 3i\gamma\Delta_p - \Delta_p^2),$$

$$a_1 = [i\gamma(2\gamma^2 + |\Omega_c|^2 - 3\Omega_s^2)\Delta_p + (2\Omega_s^2 - 3\gamma^2)\Delta_p^2 - i\gamma\Delta_p^3],$$

$$d = 4\Omega_s^2|\Omega_c|^2 + \gamma^2(\Omega_s^2 + |\Omega_c|^2),$$

$$d_0 = (|\Omega_c|^2 - \Omega_s^2)^2 + 2\gamma^2(\Omega_s^2 + |\Omega_c|^2),$$

$$d_1 = 2i\gamma[\gamma^2 + 2(\Omega_s^2 + |\Omega_c|^2)],$$

$$d_2 = 5\gamma^2 + 2(\Omega_s^2 + |\Omega_c|^2).$$

The first term in the numerator of eq. (4-a) is proportional to $p\eta e^{i\phi}|\Omega_c|$ and represents the scattering of coupling and pumping fields into the probe field via quantum interference due to the spontaneous emission. This round-trip depends on the relative phase of applied fields. Such scattering processes of the coupling and pumping fields into the probe field mode can occur even in the absence of the probe field. But in the absence of quantum interference due to the spontaneous emission, this term does not make a contribution in the probe susceptibility and consequently the response of the system to the probe field does not depend on $\phi$. The second phase-independent term in the numerator of Eq. (4-a), which is proportional to $|\Omega_p|$, represents the direct response of the medium to the probe field at the probe field frequency.

Our main observable quantity is the response of the atoms to the probe field which can be obtained via susceptibility. The susceptibility of the weak probe field is determined by the probe transition coherence $\tilde{\rho}_{31}$. In the following, we discuss the response of the atomic system to the applied field by defining the susceptibility, $\chi$, as [34]

$$\chi = \frac{2N\mu_{31}}{\varepsilon_0 E_p} \tilde{\rho}_{31} \ . \tag{5}$$

Here N is the atom number density in the medium, and $\chi = \chi' + i\,\chi''$. The real and imaginary parts of $\chi$ correspond to the dispersion and the absorption of the weak probe field, respectively. For probe transition, the transition rate and dipole moment are $\gamma = 2\pi \times 5.74\,MHz$ and $\mu_{13} = 2.53 \times 10^{-29}\,C.m$, respectively. Then the atom density $N = 5.23 \times 10^{11}\,atom/cm^3$ and probe Rabi frequency $|\Omega_p| = 0.01\gamma$, lead to $2N\mu_{13}/\varepsilon_0 E_p \cong 1$.

The group velocity of the weak probe field is then given by [34]

$$v_g = \frac{c}{1 + 2\pi\chi'(\omega_p) + 2\pi\omega_p \dfrac{\partial\chi'(\omega_p)}{\partial\omega_p}} = \frac{c}{n_g}, \tag{6}$$

where $c$ is the speed of light in the vacuum, $\chi'(\omega_p)$ is the real part of $\chi$, and $n_g$ shows the group index. Equation (6) implies that, for a negligible absorption, the group velocity can be significantly reduced via a steep positive dispersion. Moreover, the strong negative dispersion can increase the group velocity to establish even a negative group velocity.

## III. RESULTS AND DISCUSSIONS

In the first step, we ignore the effect of SGC, i.e. $\eta_1(\eta_2) = 0$ and present the numerical results of equations (2). We study the dispersion and absorption in a four-level N-type quantum system. In our notation if $\chi'' < 0$, the system exhibits gain for the probe field, while for $\chi'' > 0$, the probe field is attenuated. Figure 2 shows the dispersion (a) and absorption (b) behavior of the probe field versus probe field detuning for different values of pumping field. The selected parameters are $\gamma_{31} = \gamma_{32} = \gamma_{41} = \gamma_{42} = \gamma$, $\eta_1 = \eta_2 = 0$, $|\Omega_p| = 0.01\gamma$, $\Delta_s = \Delta_c = 0$, $|\Omega_c| = 0.12\gamma$, $\Omega_s = 0.6\gamma$ (Solid), $0.82\gamma$ (Dashed), $\gamma$ (Dotted). An investigation on figure 2 shows that for small values of pumping Rabi frequencies, the slope of dispersion is positive and the absorption doublet with an electromagnetically induced transparency (EIT) window appears in the absorption spectrum around zero probe detuning. By increasing the intensity of pumping field the slope of dispersion switches from positive to negative, corresponding to the superluminal light propagation. The peaks in absorption spectrum in figure 2(b) are explained via dressed states of atom-field Hamiltonian. Such results are in good agreement with the results of reference [29]. It is worth noting that by increasing the intensity of pumping field, the EIT absorption doublet changes to the EIT gain doublet. Then the light pulse propagates without any absorption or amplification which is an interesting situation for non-disturbed information transmission. Note that one-photon resonance condition i.e. $\Delta_s = \Delta_c = 0$ is necessary to establish the EIT window and equal decay rates assumption, i.e. $\gamma_{31} = \gamma_{32} = \gamma_{41} = \gamma_{42} = \gamma$ is necessary condition to guarantee that the changing the intensity of applied fields does not modify the EIT window.

Our analytical calculation shows that the subluminal and superluminal regions via the Rabi frequency of coupling and pumping fields, around zero probe detuning, is determined by a hyperbola relation,

$$\frac{3\Omega_s^{\,2}}{2\gamma^2} - \frac{|\Omega_c|^2}{2\gamma^2} = 1, \tag{7}$$

which is plotted in figure 3. The superluminal region is shown by dark color.

In figure 4(a), we display the group index, $\dfrac{c}{v_g} - 1$, versus probe field detuning for $\Omega_s = 0.6\gamma$ (Solid), $0.82\gamma$ (Dashed), $\gamma$ (Dotted). Other parameters are same as in figure 2. It can be realized that for $\Omega_s = 0.6\gamma$ the group index around $\Delta_p = 0$ is positive, corresponding to the subluminal light propagation. For $\Omega_s = 0.82\gamma$ the group index around $\Delta_p = 0$ is zero and the group velocity is equal to the speed of light in vacuum. By increasing the intensity of pumping field to $\Omega_s = \gamma$, the group index becomes negative corresponding to the superluminal light propagation and group velocity reaches $v_g = -1058 \ m/s$ around zero probe detuning.

Our theoretical study in this paper focuses on the analysis of the frequency transmission profile and the group velocity deduced from this profile. Such an analysis does not take into account the pulse distortion which can limit the possibility of an experimental observation the transition between slow and fast light effects. It is well known that changing in the pulse shape can be minimal in a transparent, linear anomalously dispersive medium; however, in absorption or gain lines the output pulse was slightly compressed [6, 12, 35]. Group velocity dispersion (GVD) -the frequency dependence of group velocity in a medium- has a major role in distortion of light pulse during its propagation through the transparent medium [36]. It is determined by the second derivation of susceptibility. In figure 4 (b), we show the GVD versus probe field detuning for $\Omega_s = 0.6\gamma$ (Solid), $0.82\gamma$ (Dashed), $\gamma$ (Dotted). Other parameters are same as in figure 4(a). It can be seen that the GVD is negligible around zero probe detuning.

Now, we are taking into account the effect of SGC, i.e. $\eta_1(\eta_2) \neq 0$ on the optical properties of system. In the presence of SGC, the equations (2) depend on the relative phase of applied field via quantum interference terms. Then the optical properties of the system depend on the relative phase, $\phi$. In figure 5, we show the dispersion (a) and absorption (b) of the probe field versus the probe detuning. Using parameters are $\Omega_s = 1.42\gamma$, $|\Omega_c| = 0.12\gamma$, $\Delta_s = \Delta_c = 0$, $\eta_1 = \eta_2 = 0.9$, $p_1 = p_2 = 0.7$, $\phi = 0 (Solid)$, $\pi (dashed)$. Figure 5 shows that the slope of dispersion around zero detuning for $\phi = 0$ is positive and it is accompanied by an EIT window in absorption doublet. By switching the relative phase to $\phi = \pi$, the slope of dispersion changes from positive to negative, corresponding to the superluminal light propagation. Moreover, the EIT window in absorption doublet changes to the EIT in gain doublet. Then, by changing the relative phase, the transparent subluminal light propagation switches to the transparent superluminal light propagation.

The corresponding group indexes versus the probe detuning for $\phi = 0$ (solid) and $\phi = \pi (dashed)$ are shown in figure 6(a). The other parameters are same as in figure 5. The group index for $\phi = 0$ (solid) is positive while for $\phi = \pi$ (dashed) becomes negative, corresponding to the superluminal light propagation. The related GVD versus the probe detuning for $\phi = 0$ (solid) and $\phi = \pi (dashed)$ are shown in figure 6(b). Because of negligible GVD around zero probe detuning, the probe pulse will propagate without any distortion in sub- or superluminal regions. Thus our results are more applicable only around zero probe detuning in the frequency regions with a linear dispersion which the shape of the pulse is preserved. This applies a limitation on the duration of light pulses which can propagate with reasonable distortion for a large advancement in light pulse propagation.

In the presence of SGC, the creation of EIT window depends on the intensity of applied fields. Our analytical calculations show that for $\eta_1 = \eta_2 = \eta$, $p_1 = p_2 = p$, $\gamma_{31} = \gamma_{32} = \gamma_{41} = \gamma_{42} = \gamma$ and $\Delta_s = \Delta_c = 0$, the necessary condition to establish the EIT window (absorption or gain) is that the Rabi frequency of coupling and pumping fields satisfy the following hyperbola relation,

$$\frac{\Omega_s^2}{2\gamma^2} - \frac{|\Omega_c|^2}{2\gamma^2} = 1. \qquad (8)$$

Using parameters in figure 5 and figure 6 satisfy the condition (8). If the condition (8) is not satisfied, the system may show the absorption (gain) for $\phi = 0 (\pi)$ around the zero detuning.

In order to study the propagation properties of a probe field in the medium along the propagation direction of the $z$ axis, the following Maxwell's wave equation in the slowly-varying-envelope approximation (SVEA) should be solved

$$\frac{\partial \Omega_p(z,t)}{\partial z} + \frac{1}{c}\frac{\partial \Omega_p(z,t)}{\partial t} = i\frac{\omega_p}{2c\varepsilon_0}P_p(z,t), \qquad (9)$$

where $P_p(z,t) = N\mu_{31}\tilde{\rho}_{31}$ is the macroscopic coherent polarization of the probe transition. We take the Rabi frequency of probe pulses as $\Omega_p(z,t) = \Omega_p^0 f(z,t)$ where $\Omega_p^0$ is a real maximal value of Rabi frequency and $f(z,t)$ is dimensionless normalized pulse shape function. By applying the change of variables according to the relations $\xi = z$ and $\tau = t - z/c$, the equation (9) can be written in the form

$$\frac{\partial f(\xi,\tau)}{\partial(\alpha\xi)} = i\frac{\gamma_{31}}{\Omega_p^0}\tilde{\rho}_{31}(\xi,\tau),$$  (10)

where $\alpha = \dfrac{N\omega_p\mu_{31}^2}{4\hbar\varepsilon_0 c\gamma_{31}}$. The optical Bloch equations (2) for the density matrix elements will also be the same upon substituting $z \to \xi$ and $t \to \tau$. We numerically solve the set of equations (2) and (10) on the space-time grid using finite difference method for the initial condition that the atoms are in level $|1\rangle$. It is assumed that the probe field is a Gaussian-type pulse, $f(\xi = 0,\tau) = \exp[-(Ln2)(\tau - 30)^2/\tau_p^2]$ with pulse length $\tau_p = 5/\gamma_{31}$ at the beginning of the atomic cell.

Figure 7 displays the temporal evolution of the magnitude squared of the normalized probe pulse envelope $|f(\xi,\tau)|^2$ at the entrance to the medium $\alpha\xi = 0$ (solid) and the exit of medium $\alpha\xi = 10$ (dashed) and $\alpha\xi = 20$ (dotted) for $\phi = 0(a,c), \pi(b,d)$. We put $\Delta_p = 0(a,b), 0.3\gamma(c,d)$, the other using parameters are same as in figure 5. It can clearly be shown that for $\Delta_p = 0.0(a,b)$ the probe pulse propagate with almost transparency and still preserve its shape during propagation through the medium. However, for $\Delta_p = 0.3\gamma(c,d)$ the probe pulse is attenuated (amplified) during its propagation through the medium for $\phi = 0(\pi)$ which are in good agreement with the results reported in figure 5.

The self-and cross-phase modulation due to the third order nonlinearity can affect the relative phase of applied fields during its propagation in the medium. We are interested to calculate the self- and cross-phase modulation of the probe, coupling and coherent pumping fields. The susceptibility of the probe and coupling fields are given by

$$\chi_p = \frac{N|\mu_{31}|^2}{\hbar\varepsilon_0}\frac{\tilde{\rho}_{31}}{\Omega_p} \approx \chi_p^{(1)} + \chi_p^{(3,SPM)}|E_p|^2 + \chi_p^{(3,XPM,s)}|E_s|^2 + \chi_p^{(3,XPM,c)}|E_c|^2$$

$$\chi_c = \frac{N|\mu_{32}|^2}{\hbar\varepsilon_0}\frac{\tilde{\rho}_{32}}{\Omega_c} \approx \chi_c^{(1)} + \chi_c^{(3,SPM)}|E_c|^2 + \chi_c^{(3,XPM,s)}|E_s|^2 + \chi_c^{(3,XPM,p)}|E_p|^2$$  (11)

where we have introduced the linear susceptibility $\chi_p^{(1)}(\chi_c^{(1)})$, third order self-Kerr susceptibility $\chi_p^{(3,SPM)}(\chi_c^{(3,SPM)})$ and third order cross-Kerr susceptibility $\chi_p^{(3,XPM)}(\chi_c^{(3,XPM)})$ of probe (coupling) field. According to the equations (4) the different self- and cross-phase modulation terms can be written as $\chi_p^{(3,SPM)} = \chi_p^{(3,XPM,s)} = \chi_c^{(3,SPM)} = \chi_c^{(3,XPM,p)} = 0,$

$$\chi_p^{(3,XPM,c)} = \frac{N|\mu_{31}|^2 |\mu_{32}|^2}{4\hbar^3 \varepsilon_0} \frac{i\gamma a_1}{d(d_0 + d_1\Delta_p + d_2\Delta_p^2 - 4i\gamma\Delta_p^3 + \Delta_p^4)},$$

$$\chi_c^{(3,XPM,s)} = \frac{N|\mu_{32}|^2 |\mu_{41}|^2}{4\hbar^3 \varepsilon_0} \frac{i\gamma}{d}. \tag{12}$$

Note that $\chi_c^{(3,XPM,s)}$ is pure imaginary and does not make a contribution in nonlinear phase shift during its propagation through the medium. Moreover, $\chi_p^{(3,XPM,c)}$ is also pure imaginary around zero-probe detuning and it seems that the nonlinear cross-phase shift is negligible in one-photon resonance condition of applied fields. However, for the Gaussian probe and coupling pulses, solving the propagation equations gives the nonlinear cross-phase shift $\phi_p^n$ [37]

$$\phi_p^n = \frac{2\omega_p l \hbar^2 |\Omega_c^0|^2 \, erf(\varsigma_p)}{c|\mu_{32}|^2 \varsigma_p} \mathrm{Re}\left[\chi_p^{(3,XPM,c)}\right] \tag{13}$$

where $\varsigma_p = [(1 - v_g^p/v_g^c)\sqrt{2}l]/(v_g^p \tau_c)$. The parameters $\Omega_c^0$ and $\tau_c$ show the peak Rabi frequency and time duration of coupling field, respectively. $l$ is the length of medium and $erf(\varsigma_p)$ represents the error function.

In figure 8 we plot the nonlinear phase shift of probe field induced by coupling field versus the probe detuning. Using parameters are $l = 1mm$, $\Omega_c^0 = 0.12\gamma$, $\omega_p = \omega_c = 237 \times 10^{13} Hz$, $\mu_{32} = \mu_{31}$, $\tau_c = 5/\gamma$, $\Omega_s = 0.6\gamma (solid), 0.82\gamma (dahed), \gamma (dotted)$. Other parameters are same as in figure 5. We find that the maximum nonlinear phase shift value can be controlled by the intensity of applied field. Moreover, a negligible maximum nonlinear phase shift value ($\phi_p^n = \pm 0.004\pi$) is obtained for our interested parameters.

Finally, we investigate the influence of Doppler broadening on the obtained results. The effect of Doppler broadening due to the atom's thermal velocity $\mathbf{v}$ can be included by replacing the $\Delta_p$, $\Delta_c$ and $\Delta_s$ by $\Delta_p - \mathbf{k}_p.\mathbf{v}$, $\Delta_c - \mathbf{k}_c.\mathbf{v}$ and $\Delta_s - \mathbf{k}_s.\mathbf{v}$, respectively. The parameters $\mathbf{k}_p$, $\mathbf{k}_c$ and $\mathbf{k}_s$ are the wave vector of probe, coupling and pumping fields. It is assumed that all laser fields are co-propagating in same direction, i.e. $\mathbf{k}_p = \mathbf{k}_c = \mathbf{k}_s = \mathbf{k}$. Moreover, we use a Maxwellian velocity distribution,

$$f(v) = \frac{1}{\sqrt{2\pi}D} \exp\left[-v^2/D^2\right], \tag{14}$$

to average our results for absorption, dispersion and group velocity. The parameter $D = \sqrt{2k_B T/m}$ shows the Doppler width of distribution. Then the total susceptibility of the probe field is given by

$$\tilde{\rho}_{31}(\Delta_p) = \int_{-\infty}^{\infty} f(v)\tilde{\rho}_{31}(\Delta_p, v)dv. \tag{15}$$

By including the Doppler broadening, the total susceptibility shows different behaviors for two subluminal and superluminal regions, which have been determined by equation (7) for rest atoms. Figure 9 illustrates the dispersion (solid), absorption (dashed) of probe field for different values of Doppler width, i.e. $kD = \gamma(a)$, $10\gamma(b)$, $50\gamma(c)$. Other parameters are same as the solid lines in figure 2 which satisfy the subluminal region shown in figure 3. Our numerical results show that, interestingly, by increasing the Doppler width of distribution the slope of dispersion switches from positive to negative. Then the transparent subluminal switches to the transparent superluminal light propagation, accompanied by a sub-natural structure in absorption spectrum. Recently, it was shown that the Doppler effect can induced extremely narrowed sub-natural structure [27], however such switching has not been reported. The behavior of group index versus Doppler width is shown in figure 10. When the Rabi frequency of the applied fields satisfies the subluminal region, the small values of Doppler width impose a positive group index, while by increasing the Doppler width to the room temperature ($kD \approx 50\gamma$) the group index becomes negative. Then the group velocity of light pulse can be controlled by the temperature of the medium.

The physical origin of phenomena can be understand via velocity-dependent dressed states eigenvalues which are given by [27]

$$\lambda_{c\pm} = 0.5\left(-\Delta_c + kv \pm \sqrt{(\Delta_c - kv)^2 + 4|\Omega_c|^2}\right)$$

$$\lambda_{s\pm} = 0.5\left(-\Delta_s + kv \pm \sqrt{(\Delta_s - kv)^2 + 4\Omega_s^2}\right). \tag{16}$$

According to relations (16), the position of the absorption peaks depend on the velocity of atoms. In figure 11, we display the absorption spectrum for atoms with different velocities. In left column, we plot the absorption spectrum, $\chi''(v)$, for $kv = \pm0.25\gamma(a)$, $\pm 2.5\gamma(b), \pm 5\gamma(c)$. The absorption spectrum for positive (negative) velocity is shown by dashed (dotted) lines. In right column, we show the average absorption, $[\chi''(v) + \chi''(-v)]/2$, for $kv = 0.25\gamma(d)$, $2.5\gamma(e), 5\gamma(f)$. It can be found that the atoms with small velocities impose absorption doublet and the slope of dispersion becomes positive around zero probe detuning. However, by increasing the velocity of atoms a sub-natural gain doublet is established around zero probe detuning. Thus, for small values of Doppler width, the atoms with small velocity have major contribution in the velocity averaging of optical properties. By increasing the Doppler width, the contribution of atoms with bigger velocity becomes dominant and then a gain doublet will result in the velocity averaging of absorption spectrum.

The situation is completely different for superluminal region mentioned by dark color in figure 3. Figure 12 shows the average velocity of dispersion (solid) and absorption (dashed) of probe field for different values of Doppler width, i.e. $kD = \gamma\,(a)$, $10\gamma\,(b)$, $50\gamma\,(c)$. Other parameters are same as the dotted lines in figure 2 which satisfy the superluminal region shown in figure 3. Our numerical results show that the transparent superluminal light propagation is not affected by the Doppler broadening. Moreover, by increasing the Doppler width of distribution to the room temperature, the extremely narrowed sub-natural structure is established in absorption spectrum and the slope of dispersion becomes much steeper, leading to the faster light propagation.

In figure 13, we plot the dispersion (solid) and absorption (dashed) by taking into account both SGC and Doppler effects for different values of Doppler width, i.e. $kD = \gamma\,(a,b)$, $10\gamma\,(c,d)$, $50\gamma\,(e,f)$. The left (right) column is plotted for $\phi = 0\,(\pi)$. Other parameters are same as in figure 5. An investigation on figure 13 shows that the Doppler broadening cannot disturb the phase-dependent subluminal and superluminal light propagation. However, by increasing the Doppler width, pulse distortion and absorption (gain) appears slightly in subluminal (superluminal) region.

### IV. CONCLUSION

In conclusion, the light pulse propagation was investigated in a four-level N-type quantum system. It was also shown that, in the presence of SGC, the optical properties of the system are phase-dependent and just by changing the relative phase of applied fields, the group velocity of light pulse switches from transparent subluminal to the transparent superluminal light propagation. Numerical solving of Maxwell's wave equation by finite difference approach confirmed that the probe pulse field can propagate in subluminal and superluminal regions with almost transparency and still preserve its shape during propagation through the medium. In addition, we calculated the third order self- and cross-Kerr nonlinearity and obtained a negligible nonlinear cross-phase shift of probe field for our interested system. By including the Doppler effect, it was demonstrated that increasing the Doppler width of velocity distribution to the room temperature changes the dispersion from the transparent subluminal to the transparent superluminal light propagation. Moreover, it is found that the co-propagating laser propagation directions allow us to preserve our results even in the presence of Doppler effect.

**Figure captions**

**FIGURE 1.** Schematic energy diagram of a four-level N-type atomic system driven by probe ($\Omega_p$) , coupling ($\Omega_c$) and pumping ($\Omega_s$) fields. Proposed system is established in the hyperfine levels of $^{85}Rb$ .

**FIGURE 2.** Dispersion (a) and absorption (b) behavior of the probe field versus probe field detuning for different values of pumping field. The selected parameters are $\gamma_{31} = \gamma_{32} = \gamma_{41} = \gamma_{42} = \gamma$ , $\gamma = 2\pi \times 5.74\,MHz$ , $\omega_p = 237 \times 10^{13}\,Hz$ , $\eta_1 = \eta_2 = 0$ , $\left| \Omega_p \right| = 0.01\gamma$ , $\Delta_s = \Delta_c = 0$ , $\left| \Omega_c \right| = 0.12\gamma$ , $\Omega_s = 0.6\gamma$ (Solid), $0.82\gamma$ (Dashed), $\gamma$ (Dotted).

**FIGURE 3.** Subluminal and superluminal regions via the Rabi frequency of coupling and pumping fields, around zero probe detuning. The dark color shows the superluminal region.

**FIGURE 4.** Group index, $\dfrac{c}{v_g} - 1$ (a) and group velocity dispersion (b) versus probe field detuning for $\Omega_s = 0.6\gamma$ (Solid), $0.82\gamma$ (Dashed), $\gamma$ (Dotted). Other parameters are same as in figure 2.

**FIGURE 5.** Dispersion (a) and absorption (b) of the probe field versus the probe detuning in the presence of SGC. Using parameters are $\eta_1 = \eta_2 = 0.9$ , $p_1 = p_2 = 0.7$ , $\Delta_s = \Delta_c = 0$ , $\Omega_s = 1.42\gamma$ , $\left| \Omega_c \right| = 0.12\gamma$ , $\phi = 0 (Solid), \pi (dashed)$ .

**FIGURE 6.** Group index (a) and group velocity dispersion (b) versus the probe detuning for $\phi = 0$ (solid) and $\phi = \pi (dashed)$ . Using parameters are same as in figure 5.

**FIGURE 7.** The temporal evolution of the magnitude squared of the normalized probe pulse envelope $\left| f(\xi, \tau) \right|^2$ at the entrance to the medium $\alpha\xi = 0$ (solid) and the exit of medium $\alpha\xi = 10$ (dashed) and $\alpha\xi = 20$ (dotted) for $\phi = 0 (a), \pi (b)$ . It is assumed that the probe field is a Gaussian-type pulse, $f(\xi = 0, \tau) = \exp[-(Ln2)(\tau - 30)^2 / \tau_p^{\,2}]$ with pulse length $\tau_p = 5 / \gamma_{31}$ at the beginning of the atomic cell. Other using parameters are same as in figure (5).

**FIGURE 8.** Non-linear phase shift of the probe field induced by coupling field versus the probe detuning. Using parameters are $l = 1mm$, $\Omega_c^0 = 0.12\gamma$, $\omega_p = \omega_c = 237 \times 10^{13} Hz$, $\mu_{32} = \mu_{31}$, $\tau_c = 5/\gamma$, $\Omega_s = 0.6\gamma(solid), 0.82\gamma(dahed), \gamma(dotted)$. Other parameters are same as in figure 5.

**FIGURE 9.** Dispersion (solid), absorption (dashed) of probe field versus the probe detuning for different values of Doppler width, i.e. $kD = \gamma(a)$, $10\gamma(b)$, $50\gamma(c)$. Other parameters are same as the solid lines in figure 2.

**FIGURE 10.** The behavior of group index versus Doppler width. Other using parameters are same as the solid lines in figure 2.

**FIGURE 11.** Absorption spectrum for atoms with different velocities. Left column is the absorption spectrum, $\chi''(v)$, for $kv = \pm 0.25\gamma(a)$, $\pm 2.5\gamma(b)$, $\pm 5\gamma(c)$. Right column shows the average absorption, $[\chi''(v) + \chi''(-v)]/2$, for $kv = 0.25\gamma(d)$, $2.5\gamma(e)$, $5\gamma(f)$. Other parameters are same as the solid lines in figure 2.

**FIGURE 12.** Velocity average of dispersion (solid) and absorption (dashed) of probe field versus the probe detuning for different values of Doppler width, i.e. $kD = \gamma(a)$, $10\gamma(b)$, $50\gamma(c)$. Other parameters are same as the dotted lines in figure 2.

**FIGURE 13.** Dispersion (solid) and absorption (dashed) of probe field versus the probe detuning in the presence of both SGC and Doppler effects for different values of Doppler width, i.e. $kD = \gamma(a,b)$, $10\gamma(c,d)$, $50\gamma(e,f)$. The left (right) column is plotted for $\phi = 0(\pi)$. Using parameters are same as in figure 5.

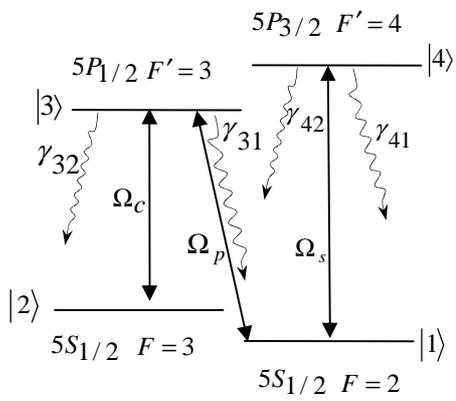

Figure 1

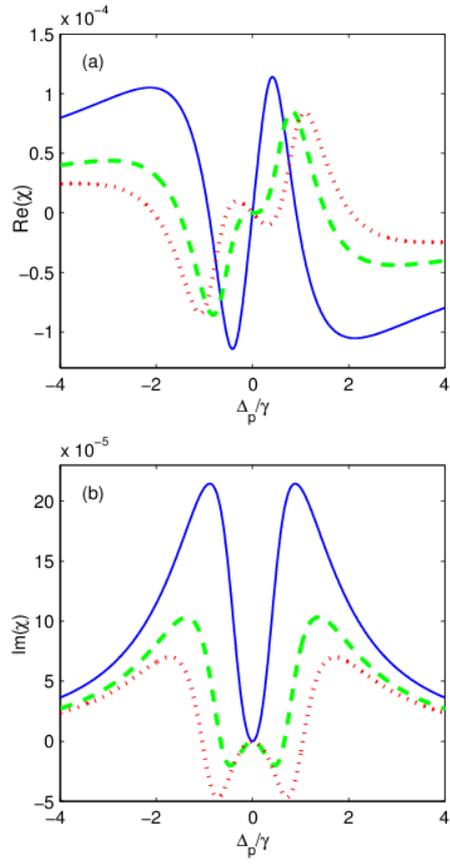

Figure 2

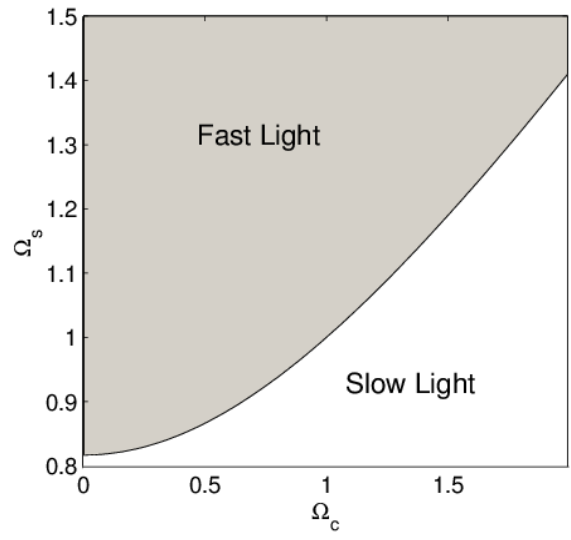

Figure 3

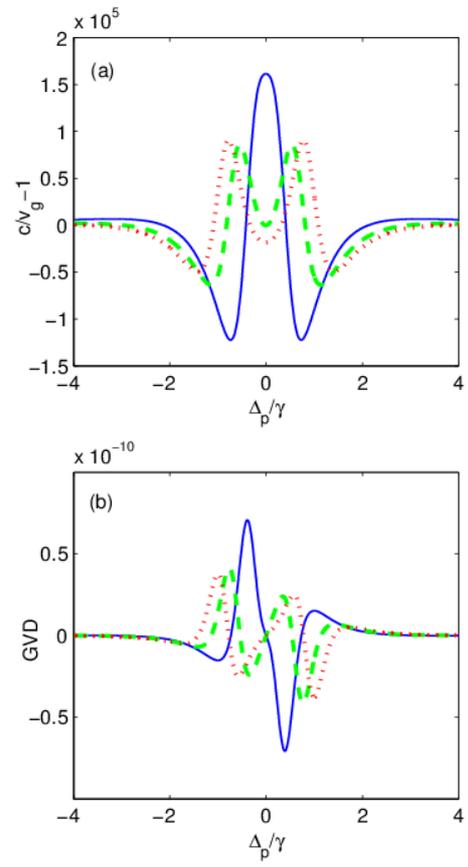

Figure 4

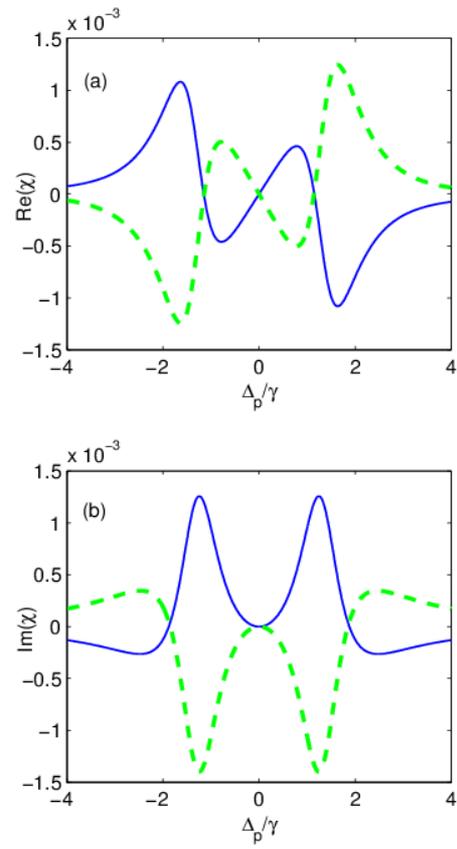

Figure 5

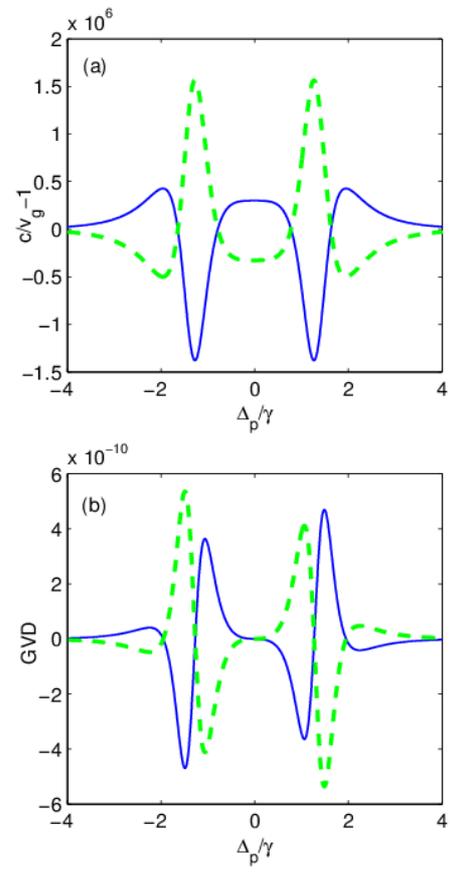

Figure 6

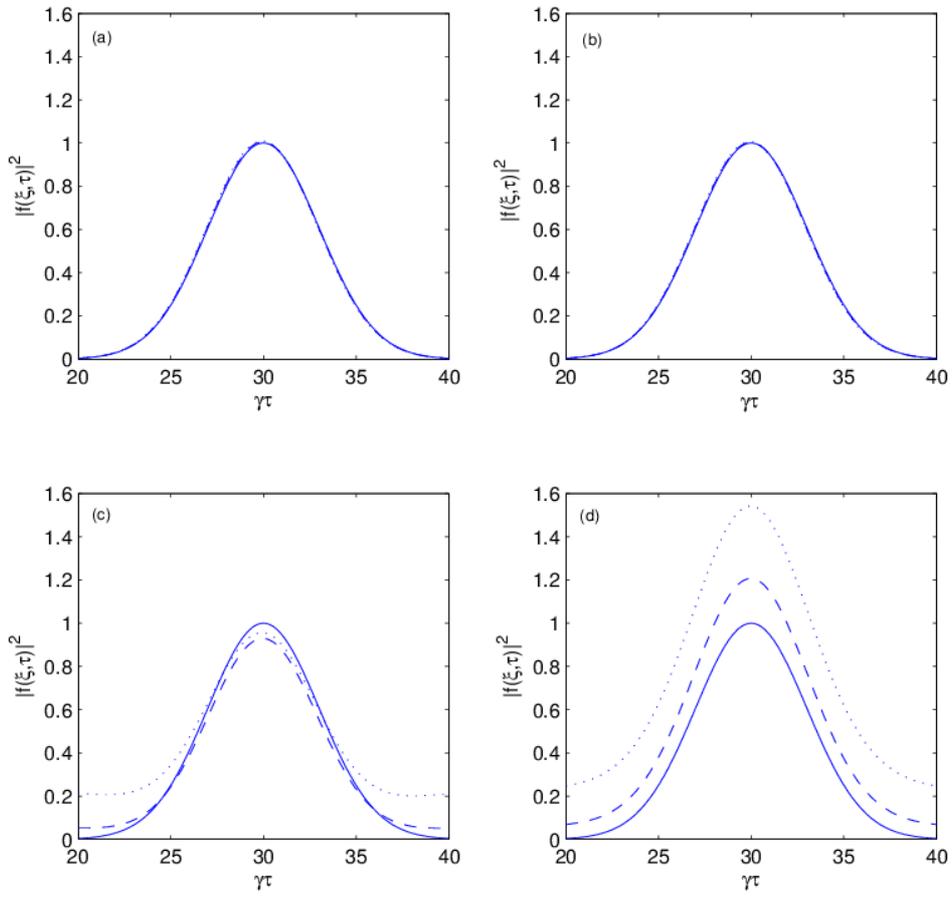

Figure 7

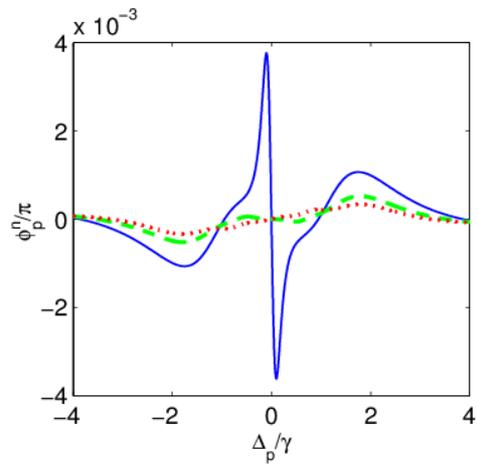

Figure 8

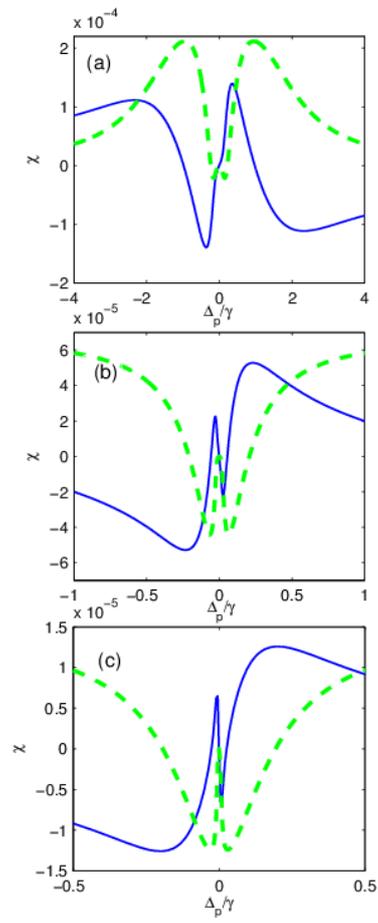

Figure 9

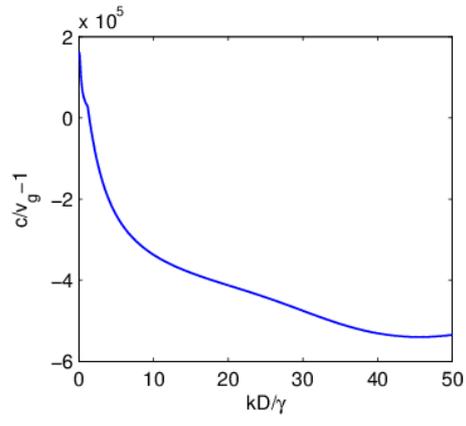

Figure 10

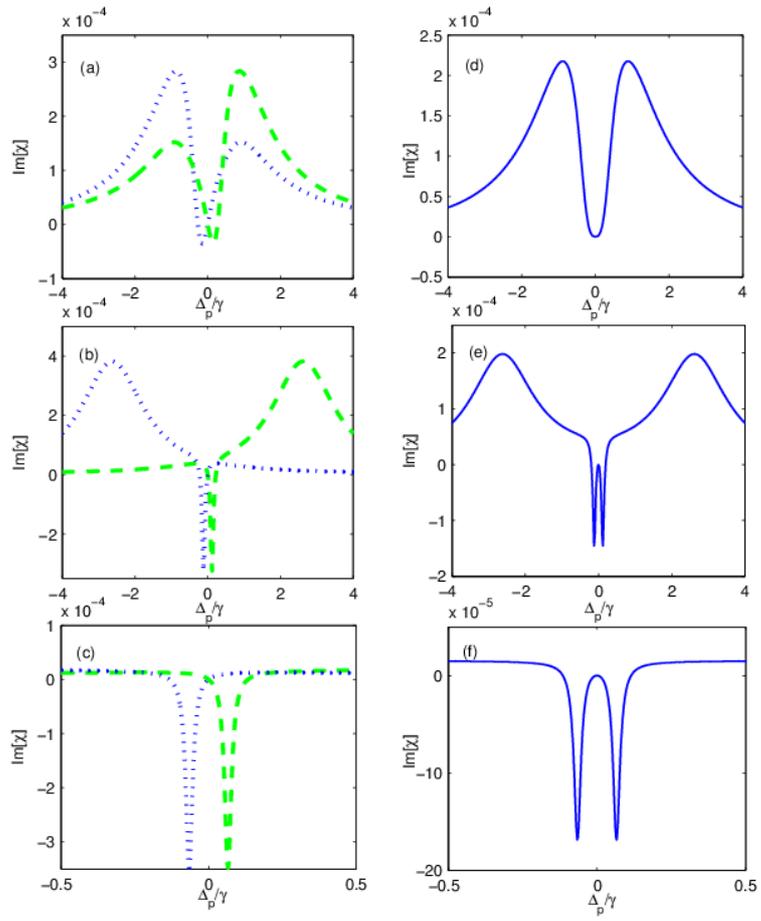

Figure 11

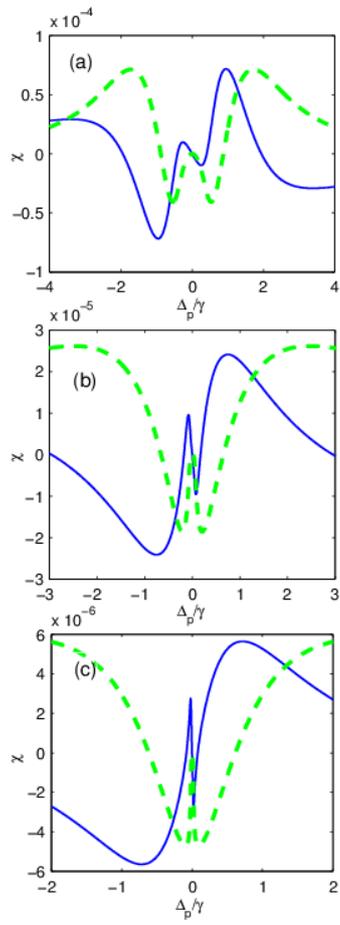

Figure 12

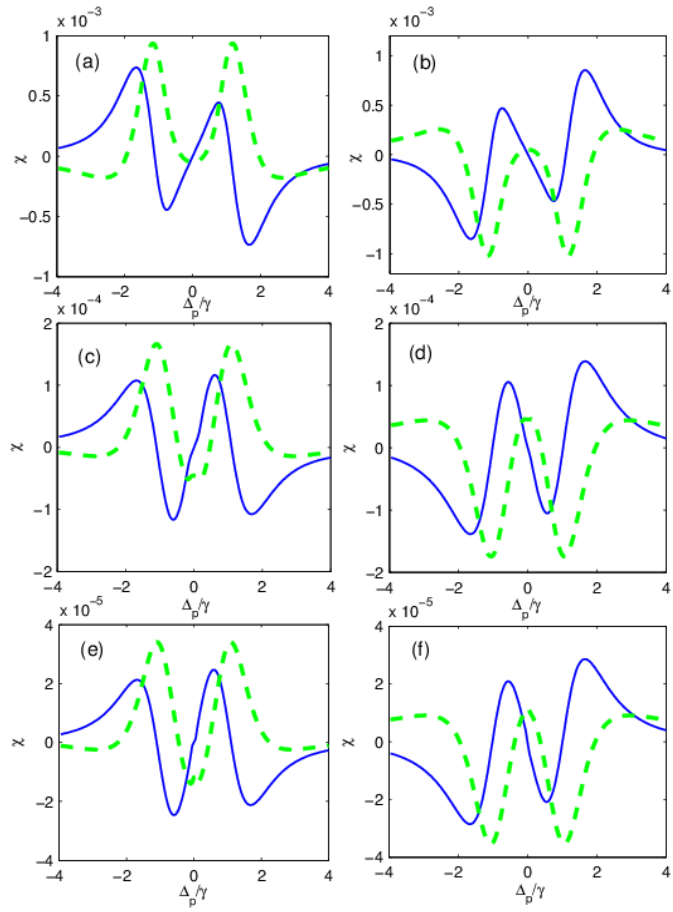

Figure 13